\documentclass[reprint,amsmath,amssymb,pra,aps]{revtex4-1}

\usepackage[T1]{fontenc}
\usepackage[utf8]{inputenc}
\usepackage{amsmath, amssymb}
\usepackage{bm}
\usepackage[dvipsnames]{xcolor}
\definecolor{dcyan}{RGB}{0,100,100}
\usepackage{graphicx}
\usepackage{tabularx}
\usepackage{braket}
\usepackage{jabbrv} 
\newenvironment{backmatter}{%
	\small%
	\newcommand{\bmsection}[1]{\par\medskip\noindent{\bfseries ##1.\enspace}}%
}{}

\newcommand{\figdir}{figures/}
\newcommand{\fignameadddir}[1]{\figdir#1}
\newcommand{\thetitle}{Photon-efficient readout of optical cavity arrays via spatial heterodyne}

\begin{document}
\title{\thetitle}
\author{Bridger McGimpsey and Matt Jaffe}

\affiliation{Department of Physics, Montana State University, Bozeman, MT}

\date{\today}

\begin{abstract}
Optical cavity arrays are an emerging tool for efficiently interfacing optical photons to quantum emitters for applications such as quantum computing. In this work, we demonstrate a metrology technique for such cavity arrays by interfering cavity array outputs against a tilted reference beam on a camera, constituting a spatial heterodyne. Both the amplitude and phase of each cavity’s transmitted field can be extracted from the resulting images. Enabled by the improved scaling of heterodyne detection over intensity detection in the presence of technical noise, we demonstrate signal-to-noise ratio greater than one down to the single detected photon level, even in the presence of hundreds of noise equivalent photons. This readout technique holds promise for fast, non-destructive imaging of atomic qubits in cavity arrays by requiring fewer scattered photons while allowing the use of fast cameras with higher electronic readout noise.
\end{abstract}

\maketitle

\section{Introduction}

Quantum emitters such as neutral atoms and trapped ions have been used for applications ranging from quantum information processing to timekeeping and sensing and are typically controlled using optical photons. Coherent exchange of quantum information between such emitters and optical photons motivates the field of cavity quantum electrodynamics (QED), in which a resonator is used to engineer strong coupling between an emitter and single optical photons. Cavity QED platforms have been used  to explore topics such as spin-squeezing~\cite{leroux_implementation_2010, cox_deterministic_2016, hosten_measurement_2016}, entanglement generation and distribution~\cite{ritter_elementary_2012, haas_entangled_2014, grinkemeyer_error-detected_2025}, analog quantum simulation~\cite{clark_observation_2020, periwal_programmable_2021, luo_hamiltonian_2025} and fundamental studies of light matter interactions~\cite{thompson_observation_1992, birnbaum_photon_2005, tanji-suzuki_interaction_2011}.

Recently the cavity array microscope has been demonstrated~\cite{shaw_cavity-array_2026, soper_stability_2026} as a cavity QED platform, in which arrays of optical resonators are constructed using a microlens array demagnified onto a neutral atom qubit array. Neutral atom qubits in a cavity array microscope hold promise for high rate entanglement distribution and efficient photon collection for fast qubit state detection. Methods for fast arbitrary addressing have also recently been demonstrated~\cite{wei_10MHz_2026, deters_ultrafast_2026}, necessitating techniques for fast atom array readout.

In this work, we present a heterodyne technique for enhanced readout of optical cavity arrays. By leveraging interference against a strong reference beam, weak optical signals can be read out with high signal-to-noise ratio (SNR). We first describe the concept behind this technique, before demonstrating both single cavity and cavity array data validating the approach. We then demonstrate the key result: high SNR down to the few-photon level robust to noisy camera readout. We then briefly outline prospects for fast, non-destructive, site-selective high-fidelity readout of neutral atom qubit arrays.

\section{Concept and approach} \label{sec:sh-concept}

\begin{figure}
	\centering
	\includegraphics[width=\columnwidth]{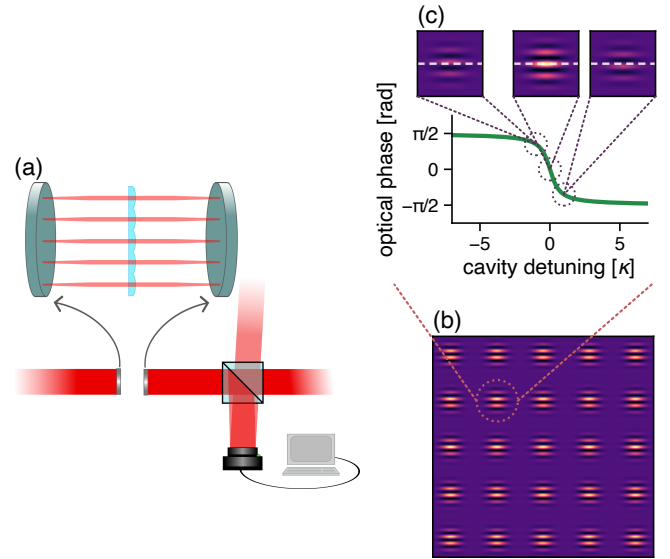}
	\caption{
		{\bf Spatial heterodyne cavity array readout concept.} (a) An array of optical cavities is formed by placing a microlens array between two flat mirrors, and excited by a global input beam. The cavity outputs are directed onto a camera along with a tilted reference beam, resulting in interference fringes shown in (b). (c) The optical phase of the cavity output field depends on the detuning relative to cavity resonance, which can then be read out via the phase (i.e., spatial position) of the interference fringes.
	}
	\label{fig:concept}
\end{figure}

The concept underlying this technique is that an interference measurement between a weak signal and a strong reference beam, the ``local oscillator'' (LO), can boost the amplitude of the weak signal above relevant technical noise sources. For a reference field $E_{\rm{LO}}$ and signal field $E_{\rm{sig}}$ populating a single optical mode, the resulting total optical field $E$ is given by $E = E_{\rm{LO}} + E_{\rm{sig}}$. When read out by a device such as a camera that measures the intensity $I = \left|E\right|^{2}$, the resulting signal can be expressed as $I = \left|E_{\rm{LO}}\right|^{2} + \left|E_{\rm{sig}}\right|^{2} + 2\left|E_{\rm{LO}}E_{\rm{sig}}\right|\cos\varphi$, where $\varphi$ is the phase difference between the two fields. For $\left|E_{\rm{LO}}\right| \gg \left|E_{\rm{sig}}\right|$, the interference term $2\left|E_{\rm{LO}}E_{\rm{sig}}\right|$ can be much larger than the raw signal intensity term $\left|E_{\rm{sig}}\right|^{2}$, not only providing a signal boost via interference with the LO, but also revealing the \emph{phase} of the signal field.

This concept can be utilized either with the LO and the signal at the same frequency or different frequencies, usually called homodyne or heterodyne, respectively. In this work, the LO and signal are at the same optical frequency, but in different spatial modes, in that the LO beam is tilted relative to signal beams, providing a constant relative phase gradient between the two fields. This is known as a ``spatial heterodyne''~\cite{kadlecek_nondestructive_2001}, moving our signal of interest to non-zero \emph{spatial} frequency where relevant noise sources are reduced compared to (spatial) dc. The same configuration is used in off-axis holography and interferometry~\cite{leith_reconstructed_1962, takeda_fourier-transform_1982, cuche_spatial_2000}, where it can reach the shot-noise limit~\cite{gross_digital_2007}, and has been applied to imaging cold atoms~\cite{kadlecek_nondestructive_2001, smits_imaging_2020} and multimode cavity emission~\cite{guo_sign-changing_2019}. Here we apply it to per-site readout of optical cavity arrays at the few-photon level.

In an ideal optical intensity measurement, the signal is given by the number of photons $N$ and shot noise is given by $\sqrt{N}$ so the signal-to-noise ratio is given by

\begin{equation} \label{eqn:snr_ideal}
	\rm{SNR}_{\rm{ideal}} = \frac{N}{\sqrt{N}} = \sqrt{N}
\end{equation}

However, in the presence of substantial technical noise $\sigma_{\rm{tech}}$ such as camera read noise, the SNR becomes

\begin{equation} \label{eqn:snr_raw}
	\rm{SNR}_{\rm{raw}} = \frac{N}{\sqrt{N + \sigma_{\rm{tech}}^{2}}}
\end{equation}

The key advantage of an interferometric technique is that instead of $N$ being represented by $ \left|E_{\rm{sig}}\right|^{2}$, the signal term of interest is the interference term $2\left|E_{\rm{LO}}E_{\rm{sig}}\right| \propto 2\sqrt{N_{\rm{LO}}N_{\rm{sig}}}$. The shot noise of the detected field now depends on $N_{\rm{LO}}$, which can be much larger than $N_{\rm{sig}}$, typically limited only by the dynamic range of the detector being used. Treating the shot noise of the signal field as signal (which it is), the SNR of the interference scheme is then

\begin{align} \label{eqn:snr_interf}
	\rm{SNR}_{\rm{interf}}  &=   \frac{2\sqrt{N_{\rm{LO}}N_{\rm{sig}}}}{\sqrt{N_{\rm{LO}} + \sigma_{\rm{tech}}^{2}}} \nonumber \\
							&=  \sqrt{N_{\rm{sig}}} \times \left(\frac{2}{\sqrt{1 + \frac{\sigma_{\rm{tech}}^{2}}{N_{\rm{LO}}}}} \right)
\end{align}

That is, for a strong enough reference beam $N_{\rm{LO}} \gtrsim \sigma_{\rm{tech}}^{2}$, we recover the SNR of the ideal measurement from Eq.~\eqref{eqn:snr_ideal}, even in the presence of substantial noise $\sigma_{\rm{tech}}$. Since $N_{\rm{LO}}$ and $\sigma_{\rm{tech}}^{2}$ both integrate over the same detector area, this condition is equivalent to a per-pixel comparison between the local oscillator fill and the read-noise variance; see Supplement 1, Sec.~\ref{app:snr-deriv} for a pixel-level treatment.

\section{Cavity characterization} \label{sec:cav-char}
In this section, we present experimental results demonstrating and characterizing the spatial heterodyne technique for cavity array readout. The experimental setup is shown in Fig.~\ref{fig:concept}. A microlens array between two flat mirrors forms an array of independent optical cavity axes; one centered on each lenslet of the array. One or many of these cavities are excited using either a mode-matched or a large, uniform excitation beam, respectively. The cavity array output is then combined on a beamsplitter with a tilted spatial heterodyne beam and imaged on a camera, generating interference fringes. These fringes encode the magnitude and phase of the transmitted electric field $E$ of the optical cavity array. The magnitude of the field $\left| E \right|$ is obtained from the fringe contrast, or the amplitude of the spatial variation. The phase is seen by the relative position of the fringes on the camera. The spatial fringe frequency is set by the angle between the beams and is straightforwardly experimentally tunable.

\subsection{Single Cavity}
We first characterize the spatial heterodyne technique using a single cavity, corresponding to one lenslet of the microlens array, excited by a mode-matched beam. To extract the transmitted cavity field $E_{\rm{cav}}$ from a camera image, the image is Fourier transformed. The interference fringes appear in Fourier space as peaks corresponding to the spatial heterodyne wavevectors $\pm k_{\mathrm{sh}}$. To extract $E_{\rm{cav}}$, we first demodulate by $k_{\mathrm{sh}}$, shifting $k_{\mathrm{sh}}$ to zero frequency. The demodulated spectrum is then Gaussian filtered and inverse Fourier transformed, resulting in a 2D array of complex values of the form $z_{ij}~=A_{ij} e^{i\phi_{ij}}$. Matched-filter averaging of $z_{ij}$ against the known cavity location and size produces an appropriately weighted average $\bar{z}~=~\bar{A} e^{i\bar{\phi}}$. The magnitude $\left|E_{\rm{cav}}\right|$ is then given by $\bar{A}$ and the phase $\arg(E_{\rm{cav}})$ is given by $\bar{\phi}$.

\begin{figure}
	\centering
	\includegraphics[width=\columnwidth]{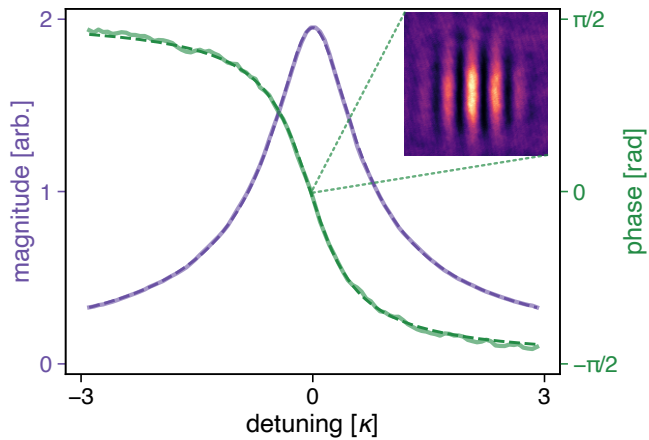}
	\caption{{\bf Single cavity characterized via spatial heterodyne readout.} Images are taken while the laser-cavity detuning is swept. The magnitude and phase of the spatial heterodyne signal directly measure the transmitted electric field. Dashed lines are fits to a square-root Lorentzian (magnitude) and an arctangent (phase). Inset: raw camera frame near resonance, showing the interference fringes from which the signal is demodulated.}
	\label{fig:single-cavity}
\end{figure}

To characterize the cavity's frequency response, we take images while sweeping the laser-cavity detuning. For each detuning, the above analysis is performed to extract the cavity output magnitude and phase. The results are shown in Fig.~\ref{fig:single-cavity}. The magnitude response follows the square root of a Lorentzian (due to the interference term $E_{\rm{cav}} E_{\rm{sh}}$, where $E_{\rm{sh}}$ is the electric field of the spatial heterodyne beam), while the transmitted cavity phase follows the expected form~\cite{siegman86}, closely approximating an arctangent near a resonance.

\subsection{Cavity Array}

The single cavity analysis can be extended to an array of cavities. Many cavities in the array are excited using a large, uniform illumination beam, as in Fig.~\ref{fig:concept}. A set of fringes is then generated at each cavity output.

\begin{figure*}
	\centering
	\includegraphics[width=\textwidth]{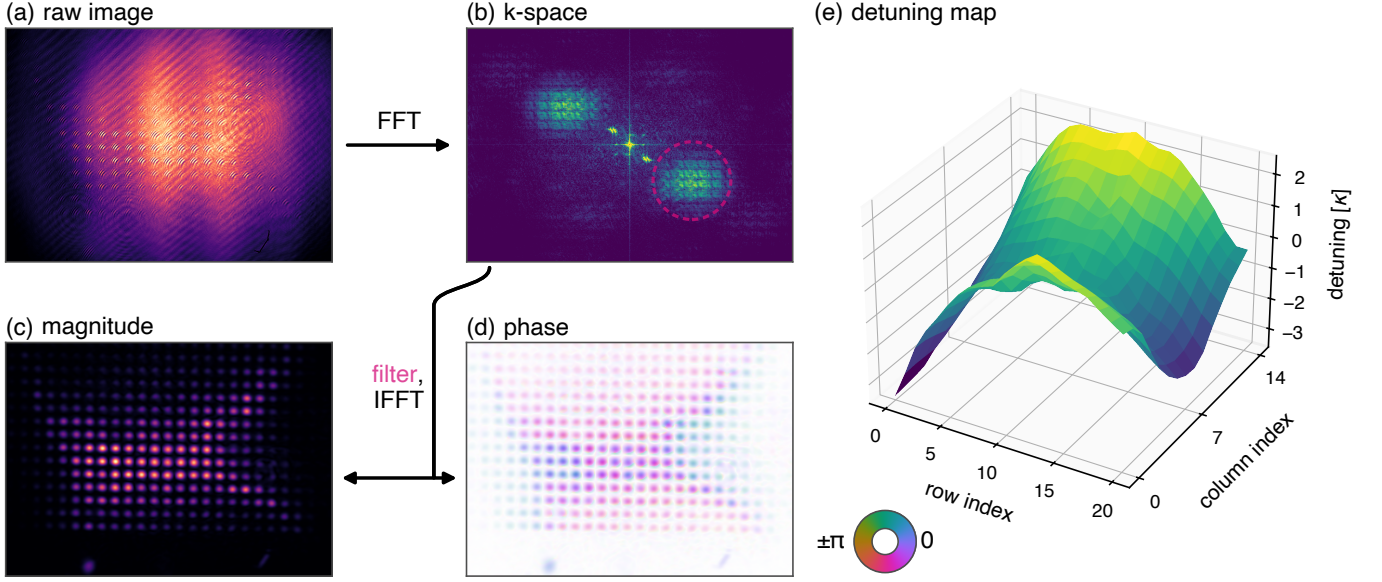}
	\caption{
		{\bf Cavity array measurement via spatial heterodyne readout.} (a) Raw image of the interference between an array of cavity modes and a large spatial heterodyne beam. (b) Fourier transform of the image reveals the spatial heterodyne component of interest (pink circle). (c, d) Demodulation shifts the spatial heterodyne component to zero frequency, before applying a Gaussian low-pass filter and inverse Fourier transform. The cavity array is clearly seen when plotting the resulting magnitudes in (c). The phases of the cavity fields extracted from the argument of the inverse Fourier transform are shown in (d) (with opacity masked by the magnitude). A static per-site phase offset due to the intentionally dirty imaging path has been subtracted in (d); see Supplement 1, Sec.~\ref{app:phase-offsets}. A reference patch for vibration/frequency shift cancellation can be seen below the cavity array. (e) A detuning map quantifying the resonance frequency homogeneity across the cavity array is extracted from the spatial heterodyne images.
        }
	\label{fig:array-data}
\end{figure*}

To extract the field of each cavity in the array simultaneously, we capture an image of the cavity array output interfering with the spatial heterodyne beam. An example image is shown in Fig.~\ref{fig:array-data}a. Note that several etalons, dust particles, and other image imperfections substantially degrade the real space image. For this dataset, these imperfections are intentionally left in order to demonstrate the robustness of the spatial heterodyne technique.

A Fourier transform is then performed. In $k$-space, the array substructure is observed, centered around the fringe frequency at $\pm k_{\mathrm{sh}}$, shown in the pink circle in Fig.~\ref{fig:array-data}b. This region of interest is demodulated to zero frequency, filtered and inverse Fourier transformed back to real space. Plotting the resulting magnitude and phase reveals the transmitted electric field of each cavity, as can be seen in Figs.~\ref{fig:array-data}c and d, respectively.

This information can be used to construct a map of the resonance frequencies across the cavity array. For cavity array applications~\cite{shaw_cavity-array_2026} (see also Sec.~\ref{sec:atom-cavity-arrays}), a uniform resonance frequency is desirable. To produce a map of the relative cavity detunings, a piezo-driven end mirror is swept while images are taken. The above spatial heterodyne analysis is performed on each image. Each individual cavity in the array can then be fitted to extract a resonance frequency (as in Fig.~\ref{fig:single-cavity}). The resonance frequency for each cavity is then plotted, resulting in the detuning map shown in Fig.~\ref{fig:array-data}e. For sufficiently low cavity resonance inhomogeneity, the same detuning map can in fact be extracted from a single frame (e.g., Fig.~\ref{fig:array-data}d) by inverting the arctan-like dependence of the transmitted cavity phase on the cavity detuning.

The detuning map provides insight into the overall cavity alignment. For example, if the cavity mirrors are not parallel, a linear detuning ramp across the array is observed. This metric can therefore provide input for high-precision alignment and characterization of more complex and alignment-sensitive cavity array geometries~\cite{shaw_cavity-array_2026, soper_stability_2026}. The astigmatic profile seen in the detuning map shown in Fig.~\ref{fig:array-data}e results from mounting strain of the thin ($t=1.2$~mm), commercial microlens array. For the cavity free spectral range $\rm{FSR}=2\pi\,\times\,$~6.9~GHz and linewidth of $\kappa=2\pi\,\times\,$~80~MHz, this corresponds to $0.01$~waves of round trip optical path length difference per linewidth of relative detuning.

\section{Low photon number detection}
Motivated by the use case of qubit detection (see Sec.~\ref{sec:atom-cavity-arrays}), we demonstrate the enhanced sensitivity of spatial heterodyne readout at low photon numbers. Detecting low photon counts with high fidelity is important for quantum emitter platforms such as neutral atom qubit arrays, where fast, low-loss readout is required. It is therefore desirable to maximize the signal-to-noise ratio at fixed, low photon number. In this section, we demonstrate a measured scaling consistent with $\sqrt{N}$ giving $\textrm{SNR}=1.3$ at a single detected photon, with SNR > 1 measured directly at 2 detected photons. These measurements are performed using only a standard CMOS camera. Low-noise cameras (e.g., EMCCD or qCMOS) can approach single-photon sensitivity in direct intensity detection, typically at the cost of frame rate or excess multiplication noise; spatial heterodyne detection instead relaxes the read-noise requirement on the camera itself. The spatial heterodyne technique outperforms a raw intensity measurement of the cavity transmission on the same camera by several orders of magnitude, as expected from the results of Sec.~\ref{sec:sh-concept}.

\begin{figure*}
	\centering
	\includegraphics[width=\textwidth]{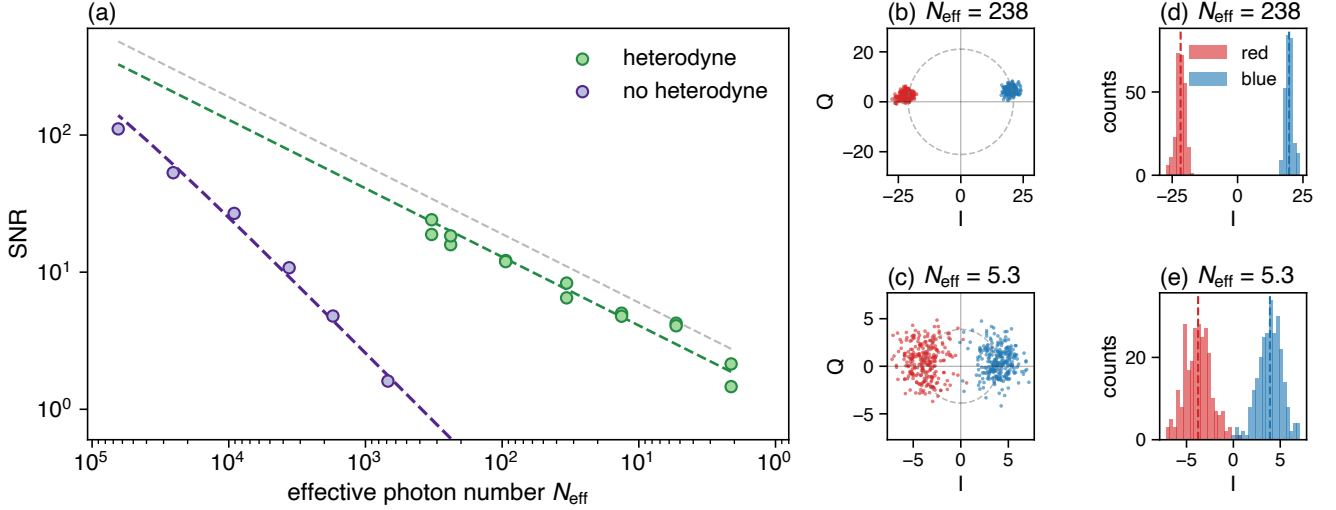}
	\caption{
	{\bf Low photon number detection using spatial heterodyne.} (a) Signal-to-noise (SNR) is plotted against photon count for both spatial heterodyne detection (green) and traditional intensity measurement (purple; i.e., with no spatial heterodyne beam). Corresponding fits are shown as dashed lines (green: $\propto\sqrt{N}$; purple: $\propto N/\sqrt{N+\sigma_{\mathrm{tech}}^2}$) and the shot-noise limit shown in gray ($\propto\sqrt{N}$). (b, c) Spatial heterodyne detection SNR is determined by plotting the demodulated signal in the IQ plane for red and blue detunings ($\pm1.5\kappa$). (d, e) Projection onto the $I$ axis is binned into histograms to further visualize separability of the two measurement conditions.
}
	\label{fig:power-data}
\end{figure*}

To simulate the use case of qubit state detection, we probe the cavity via spatial heterodyne readout at low photon numbers. The cavity is excited at detunings of $\pm~1.5~\kappa$ to induce a phase shift of $\pm\arctan(3)\approx\pm0.4\pi$ on the transmitted light. After spatially demodulating into quadratures, the resulting images can be plotted in the $IQ$ plane. SNR is then determined by analyzing the separation of point clusters for the two conditions. The resulting data is shown in Fig.~\ref{fig:power-data}.

After projecting the points on the IQ plane onto the $I$ axis, the SNR is determined via:

\begin{equation}
	\mathrm{SNR} = \frac{\left| \mu_{\mathrm{blue}} - \mu_{\mathrm{red}} \right|} {\sqrt{\sigma_{\mathrm{blue}}^2 + \sigma_{\mathrm{red}}^2}} ,
\end{equation}

where $\mu_{i}$ and $\sigma_{i}$ are the centers and widths, respectively, of the resulting distributions, with $i$ representing either the red- or blue-detuned case. The photon number incident on the camera is then varied by taking datasets with different sets of neutral density (ND) filters between the cavity output and the detection camera.
The number of photons $N_{\gamma}$ incident on the camera is determined from the transmitted cavity power $P_{\rm{trans}}$ incident on the camera with no ND filters, exposure time $t_{\rm{exp}}$, attenuation factor $\alpha_{\rm{ND}}$ of the neutral density filters, and is related to the number of photoelectrons $N_{e^{-}}$ by the camera quantum efficiency $\rm{QE}$:
\begin{equation}
	N_{\gamma} = \frac{P_{\rm{trans}}}{\hbar\omega} \times t_{\rm{exp}} \times \alpha_{\rm{ND}}  = \frac{N_{e^{-}}}{\rm{QE}},
\end{equation}
\noindent where $\hbar\omega$ is the energy of each photon. To abstract away the device-specific quantum efficiency, we use an effective photon number $N_{\rm{eff}} = N_{e^{-}}$ for our SNR analyses.

This detection scheme is compared against a raw intensity measurement, which is made via the same analysis pipeline but at spatial dc and with no interferometric enhancement. The comparison can be seen in Fig.~\ref{fig:power-data}. Note the $N^{1/2}$ scaling for the spatial heterodyne measurement, which offers advantage at low photon numbers over the $N^{1}$ intensity measurement scaling. As expected, the spatial heterodyne scheme provides signal-shot-noise-limited scaling down to the few-photon level even in the presence of significant technical noise: the fitted technical noise of the raw intensity measurement is $\sigma_{\rm{tech}} \approx 400$ effective photons. This is primarily set by the camera's per-pixel read noise: the intensity estimator integrates over the $N_{\rm{px}} \sim 4\times10^{2}$ camera pixels weighted by the extraction filter, so read noise alone ($\sigma_{\rm{cam}} \approx 9.5\,e^{-}$ per pixel) contributes $\sqrt{N_{\rm{px}}}\,\sigma_{\rm{cam}} \approx 200$ effective photons, with residual background fluctuations accounting for the remainder. The same accounting explains the robustness of the heterodyne measurement: because the local oscillator shot noise and the read noise integrate over the same pixels, the condition $N_{\rm{LO}} \gtrsim \sigma_{\rm{tech}}^{2}$ of Eq.~\eqref{eqn:snr_interf} is independent of $N_{\rm{px}}$ and reduces to a \emph{per-pixel} comparison, $N_{\rm{LO}}^{\rm{px}} \gtrsim \sigma_{\rm{cam}}^{2} \approx 90\,e^{-}$ (where $N_{\rm{LO}}^{\rm{px}}$ is the number of local oscillator photoelectrons per pixel; see Supplement 1, Sec.~\ref{app:snr-deriv}). Here the local oscillator fills each pixel with several thousand photoelectrons, leaving a read-noise penalty at the percent level.

We emphasize which features of Fig.~\ref{fig:power-data} are physically significant and which are calibration-dependent. The $\sqrt{N}$ scaling of the heterodyne data is the physical result, and is insensitive to calibration. The overall prefactor incorporates the heterodyne fringe contrast (e.g., beamsplitter polarization response and pixel-level fringe sampling) and the absolute photometric calibration of $N_{\rm{eff}}$; these multiplicative factors translate the curve without affecting its shot-noise-limited scaling.

\section{Application to qubit state detection} \label{sec:atom-cavity-arrays}

\begin{figure*}
	\centering
	\includegraphics[width=\textwidth]{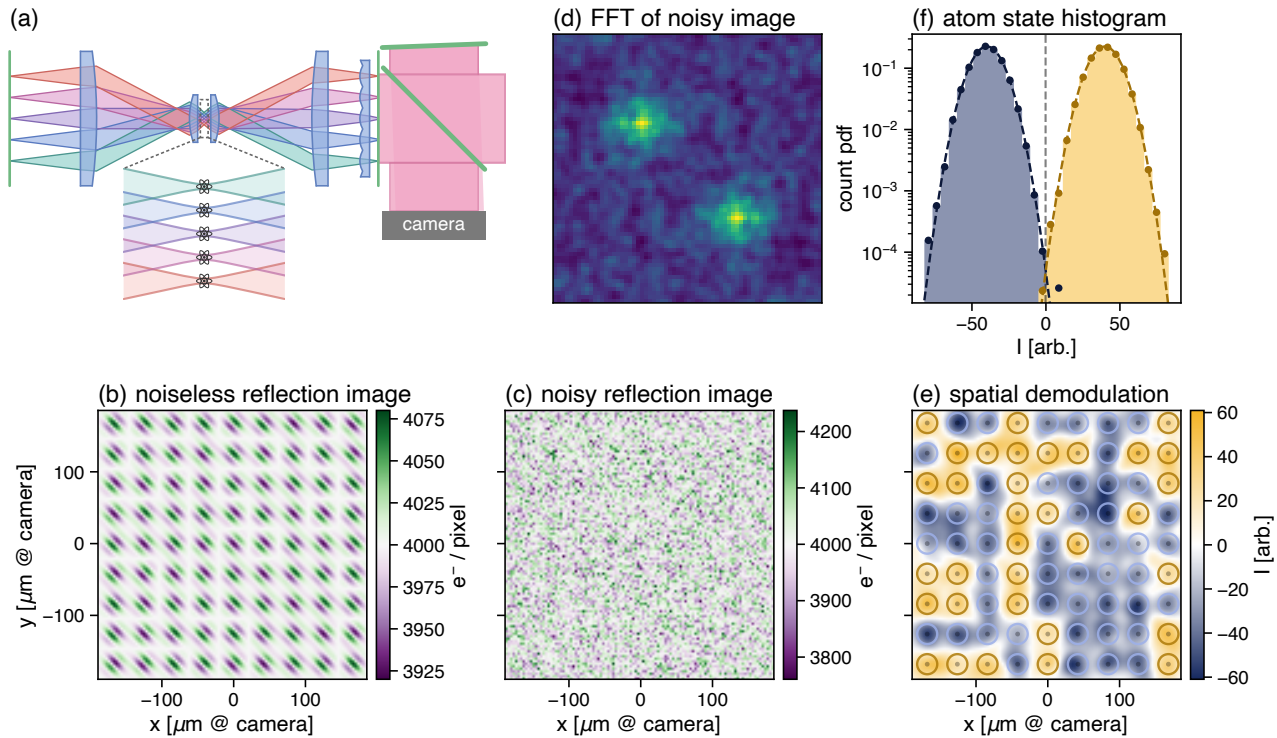}
	\caption{ {\bf Prospects for fast qubit detection (simulation).} Conceptual schematic of how spatial heterodyne detection could be used for fast, high-fidelity, non-destructive neutral atom qubit state detection. (a) A cavity array microscope with each cavity mode hosting a neutral atom qubit can be probed with a global beam to produce a per-site spatial heterodyne signal whose phase depends on the state of the qubit at that site. The resulting spatial heterodyne image (in real space) is shown noiselessly in (b) and with simulated fast camera noise in (c). (d) The Fourier transform of the noisy image reveals the signal at the spatial heterodyne frequency. (e) The state of the atomic qubit at each site is revealed via the demodulated phase. Truth values shown by the blue and gold circles around each site, indicating qubit states $\ket{0}$ and $\ket{1}$, respectively. (f) Histogramming by demodulated I quadrature component provides a high-fidelity channel for qubit state detection. Gaussian fits to the histogram are shown; dashed vertical line indicates state determination threshold. All plots shown are simulated (not experimental data) at 10 detected photons per atom.}
	\label{fig:atom-detection}
\end{figure*}

The spatial heterodyne readout technique demonstrated here may be well-suited for detection of neutral atom qubits in optical cavity arrays~\cite{shaw_cavity-array_2026}. A full detection protocol analysis is beyond the scope of this work, but the basic spatial heterodyne detection mechanism is motivated here.

State readout in neutral atom qubits is currently performed via fluorescence detection, which is slow ($\sim$~ms), global and heats the qubits, as many photons are required to be scattered by each atom to achieve the required fidelity in free space imaging systems. In a cavity of cooperativity $C$ however, Purcell enhancement induces the atom to preferentially emit into the cavity with probability $\frac{C}{C+1}$ at a rate given by $\min\left(C\Gamma, \frac{\kappa}{2}\right)$ for the atom's excited state natural linewidth $\Gamma$ and cavity linewidth $\kappa$~\cite{li_high-rate_2024}. That is, the atom can scatter photons into a \emph{single} mode faster than into free space, given sufficient cooperativity.

The spatial heterodyne technique demonstrated in this paper could enable fast, low-heating and site-selective state detection of neutral atom qubits. A rough schematic is shown in Fig.~\ref{fig:atom-detection}. Light incident on a cavity array (with each cavity hosting a neutral atom qubit) is phase-shifted depending on the state of the qubit in that cavity. The combination of preferential scattering into the cavity with spatial heterodyne-enhanced photon number sensitivity means that fewer photons need to be scattered by the atom. This is an optical analog to the temporal heterodyne readout in superconducting qubits at microwave frequencies~\cite{mallet_single-shot_2009, reed_high-fidelity_2010, walter_rapid_2017}. Spatially-selective cavity addressing (e.g., via spatial light modulator) could be used to rapidly read out all, or a subset (e.g., ancilla qubits), of the qubit array. Fast, site-selective and non-destructive imaging of this kind will be required for error correction cycles fast enough to achieve utility scale quantum computation. This spatial heterodyne detection technique opens the door to the use of high frame-rate, noisy cameras for this purpose.

\section{Conclusions}

We have demonstrated an optical spatial heterodyne technique for characterization of optical cavity arrays. The heterodyne technique successfully amplifies cavity signals, and the corresponding signal photon shot noise, well above the technical noise of even a cheap, noisy camera. Using this detection method, we have demonstrated shot noise-limited scaling down to the few-photon level. This technique will find use not only in fast characterization of optical cavity arrays, but also holds promise for fast, site-selective, non-destructive readout of neutral atom qubits, an important requirement for fault-tolerant quantum computation at utility scale.
	
\begin{backmatter}
\bmsection{Funding}
This work was supported in part by NSF award number 2514874, ARO grant W911NF-26-1-A204 and by Montana State University's Quantum Collaborative Research Environment (QCORE) through Air Force Research Laboratory agreement number FA8750-24-1-1019.

\bmsection{Acknowledgment}
We gratefully acknowledge Joshua Oliver and Saransh Bharat for their contributions to laboratory and experimental infrastructure and for helpful discussions. We also gratefully thank Andre Heinz, Mark Stone and Jon Simon for fruitful discussions about the concept and execution of the project. The U.S. Government is authorized to reproduce and distribute reprints for government purposes, notwithstanding any copyright notation thereon. Any opinions, findings, and conclusions or recommendations expressed in this material are those of the authors and do not necessarily reflect the views of the funder.

\bmsection{Disclosures}
MJ: Atom Computing (C, I), P. BM declares no conflicts of interest.

\bmsection{Data availability}
Data underlying the results presented in this paper may be obtained from the authors upon reasonable request.

\bmsection{Supplemental document}
See Supplement 1 for supporting content.
\end{backmatter}

\bibliographystyle{opticajnl}
\bibliography{spatial-heterodyne-cavity-array-bibliography}

\onecolumngrid
\vspace{1em}
\noindent\rule{0.5\linewidth}{0.5pt}
\vspace{1em}

\newpage

\begin{center}
{\large\bfseries \thetitle: supplemental document}
\end{center}
\setcounter{section}{0}
\renewcommand{\thesection}{S\arabic{section}}
\setcounter{equation}{0}
\renewcommand{\theequation}{S\arabic{equation}}
\setcounter{figure}{0}
\renewcommand{\thefigure}{S\arabic{figure}}
\setcounter{page}{1}
\renewcommand{\thepage}{S\arabic{page}}

\section{Experimental methods} \label{app:exp-methods}
The cavities probed are constructed with two partially transmissive planar mirrors with a microlens array placed between them (Thorlabs MLA300-14AR-M; lenslet pitch 300~$\mu$m, effective focal length 14.6~mm). The cavities are probed with 780~nm light from a single-mode diode laser (IPS). The cavity mode outputs are magnified and imaged onto a camera (Basler acA1280-60gm). The camera was experimentally measured to have a quantum efficiency of 27\% at 780~nm. The camera's uncoated glass window (Schott D263) reduces the number of photons arriving at the camera sensor (e2v EV76C560) by $\sim$8\%.

To lock the cavity for the low-photon number experiment, we probe two cavities in the array, a probe cavity and a lock cavity. The lock cavity is set to a desired detuning from the probe cavity by inducing a tilt in the cavity end mirrors. We then serrodyne offset lock to the lock cavity using the Pound-Drever-Hall technique, which allows us to controllably offset the relative detuning between the laser and the probe cavity. The serrodyne tone is generated by an RFSoC~4x2 FPGA board and applied with a fiber electro-optic modulator (EOSPACE). A polarization mode splitting is induced by straining a cavity end mirror with a set screw to orthogonalize the polarizations of the probe and lock cavity modes. This allows us to filter out unwanted light from the probe cavity at the lock photodetector, and from the lock cavity at the detection camera.

A reference beam, propagating along a similar path but displaced transversely from the cavity optics, also interferes with the spatial heterodyne beam. This reference interference patch is used to remove phases associated with ramping the laser frequency relative to the cavity, and to remove relative path length fluctuations of the spatial heterodyne beam path (e.g., air currents and vibrations).

\section{Single-frame detuning map} \label{app:phase-offsets}

\begin{figure*}[h]
	\centering
	\includegraphics[width=0.75\textwidth]{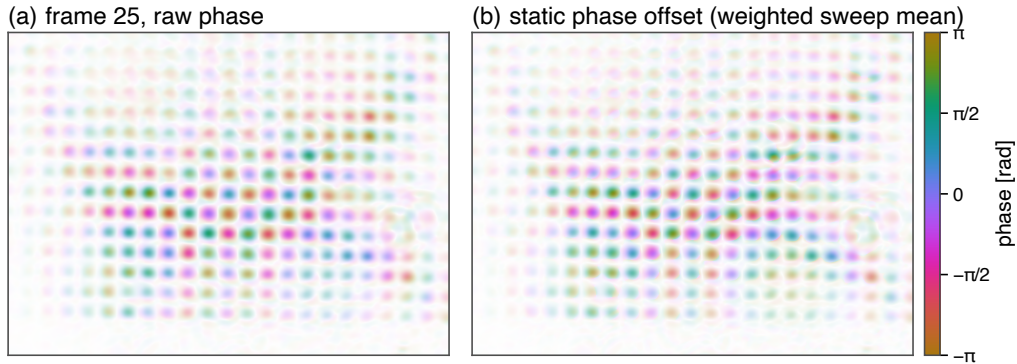}
	\caption{
		{\bf Static per-site phase offset for the cavity array of Fig.~\ref{fig:array-data}.} (a) Raw demodulated phase of the frame shown in Fig.~\ref{fig:array-data}d, before any subtraction; opacity is masked by the demodulated magnitude. (b) Static phase offset, estimated as the argument of the intensity-weighted mean of the demodulated field over the 50-frame sweep, with the same opacity mask. Fig.~\ref{fig:array-data}d is (a) minus (b).
	}
	\label{fig:static-phase}
\end{figure*}

The raw demodulated phase of the cavity array (Fig.~\ref{fig:static-phase}a) contains a static phase offset per cavity site. This offset originates outside the cavities, from wavefront disturbances of the imaging path (the etalons and dust intentionally retained for that dataset to demonstrate the robustness of the spatial heterodyne), and is static in time: the per-site phase pattern is reproduced between adjacent frames of the sweep even as the cavity resonances shift. It can therefore be measured once and subtracted. Here we estimate it per pixel as the argument of the intensity-weighted mean of the complex demodulated field over the full 50-frame sweep. The weighting makes each pixel's estimate come from the frames in which its cavity is near resonance, where the transmitted phase is near zero and the signal is far above any static stray light, while frames on either side of resonance contribute antisymmetric cavity phases that cancel. Figure~\ref{fig:static-phase} shows the raw demodulated phase of the frame in Fig.~\ref{fig:array-data}d before subtraction (a) alongside the static offset estimated this way (b). Subtracting (b) from (a) retains the cavity phase alone, which varies smoothly across the array as shown in Fig.~\ref{fig:array-data}d.

\section{Low photon number experimental details} \label{app:pi-delta}

To perform spatial heterodyne measurements at low photon number (i.e., the data in Fig.~\ref{fig:power-data}), a cavity is probed at three detunings: on-resonance and detunings of $\pm1.5\kappa$, where $\kappa$ is the cavity linewidth. This probe scheme allows us to identify low-power image ``truth'' values for their identification as ``red'' or ``blue'', and to correct for slow residual phase drifts using the phase signal from the on-resonance probe images. This measurement is performed with multiple neutral density (ND) filter stacks to vary the photon number incident on the camera.

\begin{figure*}
	\centering
	\includegraphics[width=\textwidth]{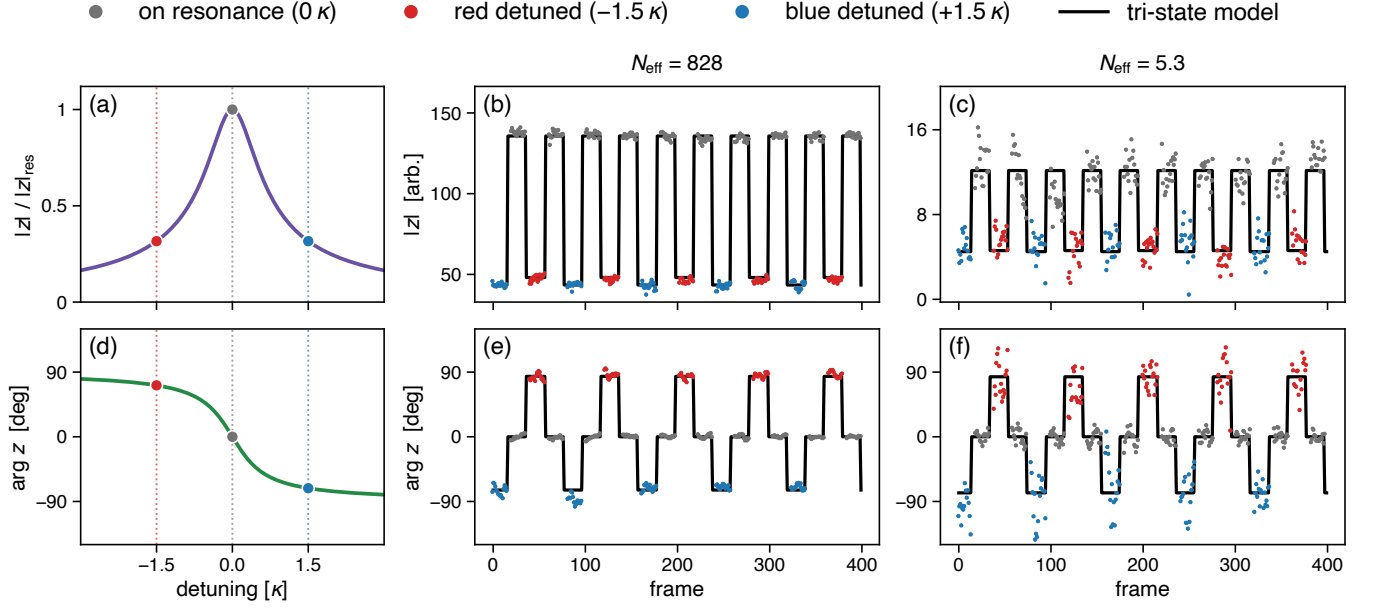}
	\caption{
		{\bf Tri-state low photon number data.} (a) Calculated cavity transmission magnitude (square root of a Lorentzian) and (d) phase (arctangent) versus laser detuning, with the three probe detunings marked: on resonance (gray) and $\pm1.5\kappa$ (red, blue). (b, c) Measured matched-filter amplitude $\left|z\right|$ and (e, f) phase versus frame for two photon numbers ($N_{\rm{eff}} = 828$ and 5.3), colored by probe state as identified from the fitted tri-state square wave (black). The on-resonance frames provide the phase reference for drift removal; the red- and blue-detuned frames form the two clusters plotted in Fig.~\ref{fig:power-data}.}
	\label{fig:tri-state-pi-delta}
\end{figure*}

To take the data underlying Fig.~\ref{fig:power-data}, the laser frequency is held via serrodyne offset lock at a sequence of detunings. At each detuning, a number of frames are taken, and the laser frequency is then rapidly swept to the next detuning. This cycle is repeated many times over a single experimental shot, with images continuously acquired at each detuning point.

These images are processed via the data analysis pipeline described in Sec.~\ref{sec:cav-char} to produce a matched-filter value for the amplitude and phase of the transmitted cavity electric field. A tri-state square wave is then fitted to this data to identify the image cases. Sample tri-state datasets can be seen in Fig.~\ref{fig:tri-state-pi-delta}.

The high-$\left|z\right|$ points are on-resonance points, while the alternating low levels are red- and blue-detuned, respectively. The on-resonance high levels allow us to identify where the red and blue-detuned data lies while also providing a local phase reference that can be used to monitor and remove slow phase fluctuations between the cavity, the reference beam, and the local oscillator. From this tri-state data, the red and blue data points are then plotted in the IQ plane (Fig.~\ref{fig:power-data}), where the $\sim0.8\pi$-phase difference between the red and blue-detuned data points is observed.

\section{Signal and noise statistics} \label{app:snr-deriv}

Here we derive the shot-noise-limited SNR of the spatial heterodyne measurement from pixel-level counting statistics, including the matched-filter demodulation and camera read noise.

Pixel $j$ of the camera records $n_{j}$ photoelectrons. The local oscillator is locally uniform with intensity $I_{\rm{LO}}$ (producing $N_{\rm{LO}}^{\rm{px}}$ photoelectrons per pixel on average) and tilt $\bm{k}_{\rm{sh}}$, so $E_{{\rm{LO}},j} = \sqrt{I_{\rm{LO}}}\, e^{i \bm{k}_{\rm{sh}} \cdot \bm{r}_{j}}$. The cavity output signal occupies a normalized mode $v_{j}$ (real, $\sum_{j} v_{j}^{2} = 1$) with $N_{\rm{sig}}$ photoelectrons and optical phase $\theta$: $E_{{\rm{sig}},j} \propto \sqrt{N_{\rm{sig}}}\, v_{j}\, e^{i\theta}$ (where the proportionality is just unit conversion constants). The mean count at each pixel is then
\begin{equation} \label{eqn:app-pixel-mean}
	\mu_{j} = N_{\rm{LO}}^{\rm{px}} + N_{\rm{sig}} v_{j}^{2} + 2\sqrt{N_{\rm{LO}}^{\rm{px}} N_{\rm{sig}}}\, v_{j} \cos(\bm{k}_{\rm{sh}} \cdot \bm{r}_{j} - \theta),
\end{equation}

\noindent with variance $\mu_{j} + \sigma_{\rm{cam}}^{2}$ (Poissonian shot noise plus Gaussian read noise), independent between pixels. The Fourier-domain demodulation and mode-weighted average described in the main text implement the matched-filter estimator

\begin{equation} \label{eqn:app-estimator}
	\hat{z} = \sum_{j} v_{j}\, e^{i \bm{k}_{\rm{sh}} \cdot \bm{r}_{j}}\, n_{j}.
\end{equation}

\noindent Terms containing residual factors $e^{i \bm{k}_{\rm{sh}} \cdot \bm{r}_{j}}$ or $e^{2 i \bm{k}_{\rm{sh}} \cdot \bm{r}_{j}}$ average to zero, as the mode $v_{j}$ varies slowly compared to the fringe period (these terms are the dc and conjugate peaks in $k$-space, removed by the demodulation filter). Writing $\cos x = (e^{ix} + e^{-ix})/2$ in Eq.~\eqref{eqn:app-pixel-mean}, the surviving term gives

\begin{equation} \label{eqn:app-mean}
	\langle \hat{z} \rangle = \sqrt{N_{\rm{LO}}^{\rm{px}} N_{\rm{sig}}}\; e^{i\theta}.
\end{equation}

For the noise, the quadratures ${\rm{Re}}\,\hat{z}$ and ${\rm{Im}}\,\hat{z}$ carry pixel weights $v_{j} \cos(\bm{k}_{\rm{sh}} \cdot \bm{r}_{j})$ and $v_{j} \sin(\bm{k}_{\rm{sh}} \cdot \bm{r}_{j})$, respectively. Since the pixels are independent with variance $\mu_{j} + \sigma_{\rm{cam}}^{2} \approx N_{\rm{LO}}^{\rm{px}} + \sigma_{\rm{cam}}^{2}$ for a strong local oscillator, and $\cos^{2}$ and $\sin^{2}$ each average to $1/2$ over the fringe while the cross term $\cos \times \sin$ averages to zero, the noise of $\hat{z}$ is circular in the complex plane, with equal, uncorrelated variance in each quadrature (and therefore along any axis):

\begin{equation} \label{eqn:app-variance}
	{\rm{Var}}\left[{\rm{Re}}\,\hat{z}\right] = {\rm{Var}}\left[{\rm{Im}}\,\hat{z}\right] = \frac{N_{\rm{LO}}^{\rm{px}} + \sigma_{\rm{cam}}^{2}}{2}.
\end{equation}

A static spatial fringe encodes both quadratures of $\hat{z}$ simultaneously: the estimator of any single quadrature, e.g. ${\rm{Re}}\,\hat{z}$, collects signal only $\propto \sum_{j} v_{j}^{2} \cos^{2}(\bm{k}_{\rm{sh}} \cdot \bm{r}_{j}) = 1/2$ of the pixels' weight, while all pixels contribute local oscillator shot noise. Spatial heterodyne detection therefore carries the standard 3~dB heterodyne penalty. A phase-locked, single-quadrature (homodyne) measurement of the two probed states, with transmitted phases $\pm\phi$ (see below), would reach $2\sqrt{2} \sin\phi \sqrt{N_{\rm{sig}}}$, a factor of $\sqrt{2}$ improvement.

In the low photon number experiments, the red- and blue- detuned states have equal mean photon number $N_{\rm{sig}}$ and transmitted phases $\theta = \mp \phi$ with $\phi = \arctan(2\delta/\kappa)$, so their means in the complex plane are separated by $\left| \langle \hat{z} \rangle_{\rm{blue}} - \langle \hat{z} \rangle_{\rm{red}} \right| = 2 \sqrt{N_{\rm{LO}}^{\rm{px}} N_{\rm{sig}}}\, \sin\phi$. With the SNR defined as this separation, projected onto the line connecting the two means, divided by $\sqrt{\sigma_{\rm{red}}^{2} + \sigma_{\rm{blue}}^{2}}$, Eq.~\eqref{eqn:app-variance} gives

\begin{equation} \label{eqn:app-snr}
	{\rm{SNR}} = \frac{2 \sin\phi \sqrt{N_{\rm{sig}}}}{\sqrt{1 + \sigma_{\rm{cam}}^{2} / N_{\rm{LO}}^{\rm{px}}}},
\end{equation}

\noindent which approaches the shot-noise limit $2 \sin\phi \sqrt{N_{\rm{sig}}}$ plotted in Fig.~\ref{fig:power-data} for $N_{\rm{LO}}^{\rm{px}} \gg \sigma_{\rm{cam}}^{2}$, independent of the local oscillator power. For the $\pm1.5\kappa$ detunings used, this corresponds to $\sin\phi\approx0.95$.

Equation~\eqref{eqn:app-snr} also makes the condition for shot-noise-limited operation precise: the local oscillator photoelectron count \emph{per pixel} must exceed the read-noise variance \emph{per pixel}, $N_{\rm{LO}}^{\rm{px}} \gtrsim \sigma_{\rm{cam}}^{2}$. For the camera used here, the measured per-pixel read noise of $\sigma_{\rm{cam}} \approx 9.5\,e^{-}$ (EMVA~1288 temporal dark noise) sets this threshold at $\sigma_{\rm{cam}}^{2} \approx 90$ photoelectrons per pixel, while the data of Fig.~\ref{fig:power-data} were taken with local oscillator fills of $>3000$ photoelectrons per pixel. The read-noise correction factor in Eq.~\eqref{eqn:app-snr} is therefore at the percent level.

\end{document}